\documentstyle [preprint,revtex]{aps}

\begin{document}
\begin{flushright}
IFP-467-UNC\\
TAR-35-UNC\\
VAND-TH-93-6\\
July 1993\\
\end{flushright}
\begin{center}
\Large
Effective Potential of a Black Hole in Thermal Equilibrium
with Quantum Fields
\end{center}
\normalsize
\begin{center}
{\large David Hochberg\footnote{hochbed@vuctrvax}}
and {\large Thomas W. Kephart\footnote{kephartt@vuctrvax}}\\
{\sl Department of Physics and Astronomy\\
Vanderbilt University, Nashville, TN 37235}\\
and\\
{\large James W. York, Jr.\footnote{york@augustus.physics.unc.edu}}\\
{\sl Institute of Field Physics and\\
Theoretical Astrophysics and Relativity Group\\
Department of Physics and Astronomy\\
University of North Carolina, Chapel Hill, NC 27599-3255}\\
\end{center}
\vfill\eject
\begin{center}
ABSTRACT
\end{center}
\begin{abstract}
Expectation values of one-loop renormalized thermal equilibrium
stress-energy tensors of free conformal scalars, spin-${1 \over 2}$
fermions and U(1) gauge fields on a Schwarzschild black hole background
are used as sources in the semi-classical Einstein equation.
The back-reaction and new equilibrium metric are solved for at
$O({\hbar})$ for each spin field. The nature of the modified black
hole spacetime is revealed through calculations of the
effective potential for null and timelike orbits.
Significant novel features affecting the motions
of both massive and massless test particles show up at
lowest order in $\epsilon= (M_{Pl}/M)^2 < 1$, where $M$
is the renormalized
black hole mass, and $M_{Pl}$ is the Planck mass. Specifically, we find
the tendency for \underline{stable} circular photon orbits, an increase
in the black hole capture cross sections, and the existence of a
gravitationally repulsive region
associated with the black hole which
is generated from the U(1) back-reaction. We also consider the back-reaction
arising from multiple fields, which will be
useful for treating a black hole
in thermal equilibrium with
field ensembles belonging to gauge theories.
\end{abstract}
\vfill\eject
\noindent
{\bf I. Introduction}

A black hole can exist in thermodynamical equilibrium provided it is
surrounded by radiation with a suitable distribution of stress-energy.
Appropriate heat baths are composed of quantum fields interacting
with the black hole geometry. The gravitational effect of the heat bath
is characterized by its gravitationally induced renormalized stress-energy
tensor. One can use the expectation value of stress-energy
tensors of quantum fields renormalized over the classical spacetime
geometry of a black hole as the source in the semi-classical
Einstein equation,
$$G^{\mu}_{\nu} = 8\pi <T^{\mu}_{\nu}>_{renormalized},\eqno(1)$$
to calculate the change induced by the stress-energy tensor on the
black hole's spacetime metric. This is the back-reaction problem
associated with the spacetime geometry of a black hole in
thermal equilibrium.

In this paper we shall use solutions of the back-reaction (1) to
investigate in detail the modifications to the Schwarzschild geometry
arising from the interaction between the black hole and various types
of quantum fields. For source terms we take the one-loop quantum
stress-energy tensors computed for conformal scalars, massless
spin-${1 \over 2}$
fermions and U(1) gauge bosons. The nature of the modified spacetime
geometry is revealed through the effective potential which completely
characterizes the motion of test particles moving in the modified
background. We calculate the effective potential for both massless
and massive test particles and separate out the back-reaction contributions
coming from each spin type $(0, {1 \over 2}, 1)$, because, as
it turns out, there
are important qualitative distinctions among the various spins.
Moreover, we will want to be able to discuss the back-reactions arising
from particular weighted sums of the separate spin-dependent cases
in order to be able to address the problem of black holes in equilibrium
with a thermal ensemble containing the field content of
gauge theories of particle physics.

{}From the properties of the renormalized stress-energy tensors we employ
and of the semi-classical Einstein equation, we can obtain accurate
fractional corrections to the metric to
$O(\epsilon)$, where $\epsilon = {\hbar} M^{-2}$,
$M_{Pl} = {\hbar}^{1/2}$ is the Planck mass
and $M$ is the mass of the black hole (Units are chosen such that
$G = c = 1$, but ${\hbar} \neq 1$ ).
This means we will be restricted to considering the $O(\epsilon)$
corrections to the effective potential. The effects we study clearly
will only be of qualitative significance for small (hot) black holes, such
as might have existed in the early universe.
Yet, already at this
lowest order, important qualitative information is contained in
the effective potential. Indeed, as we will see, there is a general
trend (for each spin case)
for the overall magnitude of the potential to decrease as the
strength of the back-reaction is increased
(i.e., as $\epsilon \rightarrow 1$) or as the number of fields increases.
For the cases we have considered, the potentials all turn over at
about $\epsilon \approx 1/2$ and tend toward local minima, but then
one is leaving the domain of validity of the calculation and
the turn-over effect is not unambiguously determined.
In particular, this
implies that the back-reaction {\it tends} to lead to the existence of
{\it stable} circular photon orbits for the null geodesics.
For the timelike geodesics, the back-reaction leads to the formation
of stable circular orbits even when the test particles have
an angular momentum {\it less} than the critical value associated
with stable orbits for the classical Schwarzschild
black hole.

Another important feature which shows up at $O(\epsilon)$
is a consequence of the gauge boson back-reaction. For this case,
and this case only, we find the back-reaction generates a repulsive
{\it anti-gravity} contribution to the net force
in the
region surrounding the renormalized black hole
event horizon. This repulsive component of the effective potential
should {\it not} be confused with the
ordinary repulsive angular-momentum-dependent barrier term
which arises in all central force problems.
Nor should it be confused with the reversal in sign of the
centripetal acceleration for a circular orbit close to an
ordinary Schwarzschild hole [1].
Indeed, the existence of this new repulsive force is confirmed
by an explicit calculation of the acceleration of a test particle
initially at rest (hence, with zero angular momentum).
In the absence of any back-reaction, the radial acceleration is
proportional to $-{\hat r}$, where ${\hat r}$ is the outward
unit radial vector. With the back-reaction, we find the
$O(\epsilon)$
correction to the
acceleration is proportional to $+{\hat r}$,
from roughly $r \approx 2.8M$ out to $r \approx 8.6M$,
indicating the presense
of an outwardly directed force. This holds for a single gauge boson
and for all $\epsilon > 0$.

The features mentioned so far reflect the physics of the back-reaction
of a single species $(N = 1)$ of each field of a
given spin. It is also interesting
to consider the effect of the back-reaction due to multiple fields
$(N > 1)$ of a given spin. This in fact is the situation one expects
to encounter when the black hole interacts with
a thermal ensemble of
fields belonging to
multiplets (representations) of specific gauge groups which arise in
modern theories of particle physics. The effective potentials for
multiple species are constructed simply by scaling the $N = 1$
potential correction terms
with
numerical coefficients which count the number, or
multiplicity, of fields of a
given spin that occur in the particular
gauge group.
We will work in the lowest approximation where all
matter and gauge fields are
free.
One consequence of this multiple field back-reaction is
that the new stable circular orbits form for smaller values of
the perturbation parameter $\epsilon$ than for the $N = 1$ cases.
Although the radius of the
corresponding domain of perturbative validity of our calculation
decreases with increasing particle
multiplicity, the new orbits all fall comfortably within these
domains for the cases we have considered.
For the U(1) case, the presence of the
repulsive ``core" and its effects become
markedly pronounced when the number of gauge bosons interacting
with the black hole exceeds a certain critical value.

\noindent
{\bf II. Stress-Energy Tensors and Solution of the Back-reaction}

Exact one-loop stress-energy tensors renormalized on a Schwarzschild
background have been computed for conformal scalar fields and for
U(1) gauge bosons, respectively, by Howard, Howard and Candelas
[2] and by Jensen and
Ottewill [3]. Both these results can be expressed in the form

$$<T^{\mu}_{\nu}>_{renormalized} = <T^{\mu}_{\nu}>_{analytic} +
 \left( {{\hbar} \over {\pi^2 (4M)^4}} \right)
\Delta^{\mu}_{\nu},\eqno(2)$$

\noindent
where the analytic piece, in the case of the conformal scalar field, was
first given by Page [4]. The term $\Delta^{\mu}_{\nu}$ is obtained from
a numerical mode sum. As this term is small in comparison to the
analytic piece, we do not include it in the calculations in this paper.
This does not affect any of our qualitative results because both pieces
seperately obey the required regularity and consistency conditions.
Moreover, the analytic piece has the correct trace anomaly in both
cases.
An analytic approximation for the stress-energy
tensor of a massless spin-${1 \over 2}$
fermion has been computed by Brown, Ottewill and Page [5]. As far as we
are aware, its accuracy has not been verified by an exact numerical
analysis, unlike the scalar and vector cases.

Each of the above mentioned tensors has the asymptotic form of a flat
spacetime radiation stress-energy tensor at the uncorrected
Hawking temperature ($T_H$)
at infinity of an ordinary Schwarzschild
black hole:
$$ T^{\mu}_{\nu} \rightarrow  {\rm const.} \times
{\rm diag}(-3,1,1,1)^{\mu}_{\nu}.\eqno(3)$$
In what
follows, it is convenient to write
the radial and lateral pressures of the $U(1)$ radiation at infinity
as (no sum on $i$)

$$(T^i_i)_{\infty} \equiv {1 \over 3}a T_H^4
= {{\epsilon} \over {48 \pi K M^2}},\eqno(4)$$
where $K = 3840 \pi$, $a =({\pi}^2 / {15 {\hbar}^3})$,
$T_H = {\hbar}/(8\pi M)$ is the Hawking temperature at infinity,
and $i = r,\, \theta,$ or $\phi$.
The analytic expressions all satisfy
${\hat \nabla}_{\mu} T^{\mu}_{\nu} = 0$
on the Schwarzschild background.

The back-reaction induced by the scalar, spinor and vector fields
is solved by calculating the
fractional corrections $h^{\alpha}_{\nu}$ to the metric, obtained by
setting
$$g_{\mu \nu} = {\hat g}_{\alpha \mu} [\delta^{\alpha}_{\nu} +
 \epsilon\, h^{\alpha}_{\nu}]\eqno(5)$$
in the semi-classical Einstein equation (1). We work in linear order
in $\epsilon$ as required by ${\hat \nabla}_{\mu} T^{\mu}_{\nu} = 0$
and ${\hat \nabla}_{\mu} (\delta G^{\mu}_{\nu}) = 0$, where
$\delta G^{\mu}_{\nu}$
is the Einstein tensor linearized on a background satisfying
${\hat G}^{\mu}_{\nu} = 0$.
The corrected geometry will be taken to be
static and spherically symmetric. Working out the equations as in [6],
we find the corrected metric can be written as
$$ds^2 = -\left(1 - {{2m(r)} \over r} \right)
 \left(1 + 2 \epsilon {\bar \rho}(r) \right) dt^2
 + \left(1 - {{2m(r)} \over r} \right)^{-1} \!dr^2 + r^2 d{\Omega}^2,
\eqno(6)$$
where $d\Omega^2$ is the standard metric of a normal round unit sphere.
To obtain $m(r)$ and ${\bar \rho}(r)$ requires only simple radial
integrals involving $T^t_t$ and $T^r_r$. The angular components enter
linearized Einstein equations that hold automatically by
virtue of ${\hat \nabla}_{\mu} T^{\mu}_{\nu} = 0$ in a static
spherical geometry.

The mass function $m(r)$ has the form
$$m(r) = M(1 + \epsilon\,\mu(r) + \epsilon\, C K^{-1}),\eqno(7)$$
with
$$\mu(r) = {1 \over {\epsilon M}}\, \int_{2M}^{r} (-T^t_t)\,
4\pi{\tilde r}^2 \, d{\tilde r},\eqno(8)$$
and $C$ is an integration constant which serves to renormalize
the bare Schwarzschild mass $M$, as discussed in [7].
The metric is completed by a determination of ${\bar \rho}$ which, like
$\mu$, can be found from an elementary integration. Defining
$$K\,{\bar \rho} \equiv K\,\rho + k,\eqno(9)$$
where $k$ is a constant of integration, we have
$$\rho = {1 \over {\epsilon}} \int_{2M}^{r} (T^r_r - T^t_t)
({\tilde r} - 2M)^{-1} 4\pi{\tilde r}^2 \, d{\tilde r}.\eqno(10)$$

Denote with a subscript $``s", ``f"$ and $``v"$ the metric functions
and corresponding integration constants connected with the scalar, fermion
and vector back-reactions, respectively. Then, substituting the
relevant components of the
corresponding stress-energy tensors from [2-5]
into
the formulas for $\mu(r)$ and $\rho(r)$ yields (with $w \equiv {2M}/r$)

$$K\,\mu_s = {1 \over 2}({2 \over 3}w^{-3} + 2w^{-2}
+ 6w^{-1} - 8\ln(w) - 10w - 6w^2 + 22w^3 - {{44} \over 3}).\eqno(11)$$

$$K \rho_s = {1 \over 2} \left( {2 \over 3}w^{-2} + 4w^{-1} -
8\ln(w) - {{40} \over 3}w - 10w^2 - {{28} \over 3}w^3 + {{84} \over 3}
\right).\eqno(12)$$

for the conformal scalar field [6],

$$K\mu_f = {7 \over 8} \left({2 \over 3}w^{-3} + 2w^{-2} + 6w^{-1}
- 8\ln(w) - {90 \over 7}w -{62 \over 7}w^2 + {46 \over 3}w^3 - {16
\over 7} \right), \eqno(13)$$

$$K\rho_f = {7 \over 8} \left({2 \over 3}w^{-2} + 4w^{-1}
- 8\ln(w) - {200 \over 21}w -{50 \over 7}w^2 - {52 \over 7}w^3 + {136
\over 7} \right), \eqno(14)$$

for the massless spinor [7], and

$$K\,\mu_v = {2 \over 3}w^{-3} + 2w^{-2} + 6w^{-1} - 8 \ln(w)
 + 210w - 26w^2 + {{166} \over 3}w^3 - 248.\eqno(15)$$

$$K\,\rho_v = {2 \over 3}w^{-2} + 4w^{-1} - 8\ln(w) +
{{40} \over 3}w + 10w^2 + 4w^3 - 32,\eqno(16)$$

\noindent
for the U(1) vector field [8].
Note that at the horizon $r = 2M$, or $w = 1$, we have
$\rho_s(1) = \rho_f(1) = \rho_v(1) = 0$.
The constant $k$ for the scalar, spinor and vector
is denoted by
$k_s, k_f$ and $k_v$, respectively, and will be determined
below by a boundary condition.

As pointed out in [6], the back-reaction problem (1) has no definite
solution unless boundary conditions are specified at a certain radius
$r_o$. Moreover, (3) indicates that the stress-energy tensors are
asymptotically constant, thus the combined system of the black hole
plus equilibrium quantum fields must be put into a finite ``box" or
cavity [7]. This is to insure that the fractional corrections
$\epsilon h^{\alpha}_{\nu}$ to the metric remain small for sufficiently
large radius. Obviously, the ``box" is merely a device to provide
reasonable boundary conditions that mimic implantation of the hole into
the universe. Under suitable conditions, as discussed in [7], the
specific boundary conditions should not affect our results significantly.
In what follows, we shall assume that the cavity radius
is sufficiently large that the stress-energy tensors we employ, which were
computed for infinite space, are a good approximation. If the radius $r_o$
were to approach the horizon, explicit size and boundary effects would
have to be taken into account in the construction of $<T_{\mu \nu}>$, as
shown in [9,10].

One convenient way to fix the constants $k_s$, $k_f$
and $k_v$ is to
impose a microcanonical boundary condition [6].
We fix $r_o$ and imagine placing there an ideal massless perfectly
reflecting wall. Outside $r_o$, we then have an ordinary
Schwarzschild spacetime
$$ds^2 = - \left( 1 - { {2m(r_o)} \over r} \right) dt^2 +
\left( 1 - { {2m(r_o)} \over r} \right)^{-1} \! dr^2 + r^2 d\Omega^2,
\eqno(17)$$
for $r \geq r_o$.
Continuity of the three-metric induced by metrics (6) and (17)
on the world tube $r = r_o$ fixes the constant $k$, i.e., $k_s$
$k_f$
or $k_v$, in $\bar \rho$ by the relation
$$k = - K\, \rho(r_o).\eqno(18)$$
There are finite discontinuities in the extrinsic curvature
of the world tube
$r = r_o$ [6], but these, and other properties of the box
wall, are of no interest in the present analysis, as argued in [7].
The spacetime
geometry, including back-reaction, is now completely determined by
(17) for $r \geq r_o$, and for $r \leq r_o$ by
$$ds^2 = -\left( 1 - { {2m(r)} \over r} \right)
[1 + 2\epsilon\,(\rho(r) - \rho(r_o)) ] dt^2 +
\left( 1 - { {2m(r)} \over r} \right)^{-1} \! dr^2 + r^2 \,d\Omega^2
.\eqno(19)$$ We note the metric is continuous at the boundary
$r = r_o$.

As mentioned above, the choice of box radius $r_o$
must be made so that the corrections to the
background metric remain uniformly small.
In other words, we must
establish the domain of validity implicit in regarding the effect of
$T^{\mu}_{\nu}$ as a perturbation of the Schwarzschild geometry. This
requirement will be met provided that
$$|{{\Delta g_{\mu \nu}} \over {\hat g_{\mu \nu}}}|
\equiv \delta < 1,\eqno(20)$$
where the metric perturbations, $\Delta g_{\mu \nu} = g_{\mu \nu} -
{\hat g_{\mu \nu}}$, are given by
$$\Delta g_{tt} = -2\epsilon\, {\bar \rho(r)}
\left( 1 - {{2M} \over r} \right) +
  \epsilon\, {{2M} \over r}\, \mu(r), \eqno(21)$$
and
$$\Delta g_{rr} = \epsilon\, {{2M} \over r}\,\mu(r)
\left(1 - {{2M} \over r} \right)^{-2}.
\eqno(22)$$
The angular components receive no corrections.
Both perturbations have identical asymptotic magnitudes:
$$\lim_{r \rightarrow \infty}\,|{{\Delta g_{\mu \nu}} \over {\hat g_{\mu
\nu}}}|
 = \epsilon\, \alpha_{j}
\left({2 \over {3K}} \right) \left({r \over {2M}}\right)^2,\eqno(23)$$
where $\alpha_{s} = {1 \over 2},\, \alpha_{f} = {7 \over 8}$,
and $\alpha_v = 1$, for
the scalar, spinor and vector cases, respectively. The domain radius,
$r_{dom}$, is defined to be the upper bound for which the metric
perturbations (21,22) are uniformly small:
$$1 \leq \left({ {r_{dom}} \over {2M}}\right)^2 = {{3K} \over {2 \alpha_j}}\,
 \left({{\delta} \over {\epsilon}}\right), \eqno(24) $$
obtained by substituting the asymptotic forms (21) and (22) into (20).
Thus, one must take $\delta < 1$ and
$r_o \leq r_{dom}$ to ensure the validity of the
perturbative calculations.

\noindent
{\bf III. Effective Potential}

To explore the potential in the vicinity of the black hole, we can,
without loss of generality, consider an equatorial slice
$\theta = \pi/2$ of the corrected geometry (19). Then the four-velocity
of a test particle in that background is
$$U^{\mu} = ({\dot t}, {\dot r}, 0, {\dot \phi}),\eqno(25)$$
where the overdot denotes differentiation with respect to either
the proper time or an affine parameter, depending on whether the
test particle is massive or massless. The square of this four-velocity
is
$$g_{\mu \nu}U^{\mu}U^{\nu} = g_{tt}{\dot t}^2 + g_{rr}{\dot r}^2 +
g_{\phi \phi}{\dot \phi}^2 = -\kappa,\eqno(26)$$
where $\kappa = 0$ or $1$, for the null and timelike cases, respectively.
Because the modified spacetime geometry (19) is static and spherically
symmetric, there exist two conserved quantities corresponding to the two
Killing vectors
$(\partial/{\partial t})^{\nu} \equiv (1,0,0,0)^{\nu}$ and
$(\partial/{\partial \phi})^{\nu} \equiv (0,0,0,1)^{\nu}$
of this geometry. As for the
Schwarzschild case, these constants of the motion
are identified with the
particle's total energy, $E$, and orbital angular momentum
 (with respect to the origin), $L$:
$$E = - g_{\mu \nu} ({{\partial} \over {\partial t}})^{\mu} U^{\nu} =
 - g_{tt}\, {\dot t},\eqno(27)$$
$$L = g_{\mu \nu} ({{\partial} \over {\partial \phi}})^{\mu} U^{\nu} =
 g_{\phi \phi}\, {\dot \phi} = r^2\,{\dot \phi}.\eqno(28)$$

Combining these with (26) and using the metric components from (19) yields
the test particle's geodesic equation,

$$[1 + 2\epsilon {\bar \rho(r)}] {\dot r}^2 +
\left(1 - {{2m(r)} \over r}\right)
[1 + 2\epsilon {\bar \rho(r)}]\left(\kappa + {{L^2} \over
{r^2}}\right) = E^2. \eqno(29)$$
The dependence of (29) on the boundary constant $\rho(r_o)$ is related to
the calibration of coordinate time ($t$) versus proper time ($\tau$).
Since coordinate time has no special meaning unless the metric is
asymptotically constant, we might have chosen the timelike Killing
vector to be $({{\partial} \over {\partial {\bar t}}}) = \lambda
({{\partial} \over {\partial t}})$ for $\lambda = {\rm const.}$,
instead of $({{\partial} \over {\partial t}})$. This corresponds to the
rescaling $E \rightarrow {\bar E} \equiv \lambda E$. Deriving the
geodesic equation with this choice and taking $\lambda = [1 +
\epsilon \rho(r_o) ]$ yields (29) with $E^2$ replaced by ${\bar E}^2$
and ${\bar \rho}$ replaced by $\rho$, i.e., the integration constant
$\rho(r_o)$ has been absorbed into the total energy. Note, however, that
for $r \rightarrow \infty$, ${\bar E}$ does not reduce to the special
relativistic formula ($ = {dt} /{d\tau}$) for the total energy
(per unit rest mass) of a particle (as seen by a static observer)
unless $\lambda = 1$. Although it is significant that results come out
without reference to a possibly non-existent flat distant region, the
direct comparison with the standard Schwarzschild case may be helpful, and
we shall use $E$, not ${\bar E}$, in what follows. No real physical
observable should depend on whether one uses $E$ or ${\bar E}$.

For circular orbits, ${\dot r} = 0$, and the total energy $E$ of the
particle
is just a function of $r$ (velocity-independent).
Thus, the effective potential for a black hole
modified by $O({\hbar})$ stress-energy is
$$V(w) = \left(\kappa + {{L^2} \over {4M^2}} w^2 \right)\, (1 - w)
\left[ 1 + \epsilon(2{\bar \rho(w)} - w(1 - w)^{-1}\,\mu(w)) \right].
\eqno(30)$$
Equation (29) is a differential equation for the radial coordinate.
Once the radial motion is determined using this effective potential, the
time coordinate change (relevant only if one refers to time at flat spatial
infinity)
and angular motion are easily found from
(27) and (28).
For $\epsilon = 0$, $V$ reduces to
the effective potential of a classical
Schwarzschild black hole [11]. The function defined
in (30) plays the role of an effective potential, in the
sense that the condition $E^2 > V$ determines the classically admissable
range of the point particle's motion.

\noindent
{\bf A. Null Orbits}

Setting $\kappa = 0$ and substituting the appropriate functions
$\mu$ and $\rho$ from the solutions (11-16)
of the back-reaction into (30)
yields the effective potential for null geodesics arising from
scalar, fermion or gauge boson back-reactions. For simplicity and
purposes of illustration, we set $\delta = \epsilon$ in (24) and
find the domain radii to be $r_{dom} = 380M,\, 286M$ and $268M$ for
spin $0, {1 \over 2}$ and $1$. Since the shape of $V$ is independent
of the angular momentum $L$, we plot in Fig. 1 the
functions $(4M^2/L^2)\,V$, using the appropriate
functional forms for $\mu(w)$ and $\rho(w)$, to indicate
the nature of the back-reaction for
the different spin cases and for various choices of $\epsilon$
$(\epsilon = 0.,\, .15,\, .3,\, .45,\, .6)$.
The $\epsilon = 0$ (i.e., no back-reaction) case is displayed for reference.
The effective potentials corresponding to the single-particle
back-reactions are qualitatively indistinguishable among the
scalar, spinor or vector cases, and Fig. 1 shows  $(4M^2/L^2)\,V$
for the conformal scalar.
We note for $\epsilon \stackrel{<}{\sim} 1/2$, the effective
potential is qualitatively similar to the Schwarzschild case
(i.e., no back-reaction) in that they exhibit maxima corresponding
to a single unstable circular photon orbit, and no local minima.
The location of this
maximum follows from solving the equation ${\partial V}/{\partial w} = 0$
or
$$w(2 - 3w) + \epsilon \left[ 2w(2 - 3w) {\bar \rho}_j - 3w^2 \mu_j
 - {{32\pi M^2} \over {\epsilon \, w}}\, (T^r_r)_j \right] = 0,\eqno(31)$$
for $j = s,f$ or $v$. Remarkably, the position
of the unstable circular
orbit is
relatively independent
of variations in $\epsilon$, and
we find that ${\bar r} = 3.0M$ solves (31)
for all spin cases to a very good approximation.
What does depend strongly on $\epsilon$ is the overall magnitude of the
potential. This tends to decrease as $\epsilon$ increases.
Eventually, as perturbation theory becomes unreliable
(for $\epsilon \stackrel{>}{\sim} 1/2$), the potentials tend to turn over.
This implies the
tendency for the potentials to develop
{\it stable} circular
photon orbits, a novel feature which is absent for the
Schwarzschild black hole. Since the potential is
becoming negative in this
instance, these orbits will exist as bound states $(E < 0)$.

The lowering of the potential barrier, which occurs well within the
perturbative region,
profoundly affects
the ability of the black hole to capture photons (and neutrinos, gravitons,
or any massless quanta).
The minimum energy $E$ required to surmount the top of the potential
barrier is $E({\bar r}) = V({\bar r})$.
The solutions of this equation
represent the (classical) turning points of the effective potential.
The apparent impact parameter
of a light ray, i.e., the distance of closest approach to $r = 0$, is
$b = L/E$, and the black hole will capture any light ray sent towards it
whose impact parameter is less than the critical value:
$$b_{crit} \equiv {{L} \over {E({\bar r})}}.\eqno(32)$$
Thus, the photon capture cross section for the equilibrium black hole is
$$\sigma_{capture} = \pi b_{crit}^2 \geq 27\pi M^2,\eqno(33)$$
and is larger than the Schwarzschild value of $27\pi M^2$ [12].

\noindent
{\bf B. Timelike Orbits}

When $\kappa = 1$ the shape of the effective potential depends
on the test particle's angular momentum in an important way.
Recall, for the case of the Schwarzschild black hole, $V$ will
have no extrema if $L < L_{crit} = 2\sqrt{3} M$, and a particle
heading towards the center of attraction will fall into the
singularity no matter how far away it is initially. By contrast, when
$L > L_{crit}$, the effective potential has a maximum
and a local minimum,
associated with unstable and stable
circular massive orbits. Furthermore, when
$L > 4M$, the potential will have massive bound orbits [11].
As we will see, new
features show up due to the back-reaction.

Calculations of $V$ are summarized graphically in Figures 2-4, where
we have plotted the effective potential for various values of $L$ and
$\epsilon$ for the case of the spin-${1 \over 2}$ back-reaction
(the other cases are qualitatively
similar in every respect and we do not include them here).
The no-back-reaction curve is included for reference.
When $L < L_{crit}$, $V$ has no local extrema as shown in Fig. 2, but the
magnitude of $V$ decreases
for increasing $\epsilon$, just as for the null-orbit
examples discussed above.
The potential changes sign around $\epsilon \approx
1/2$. In Fig. 3 the effect of higher angular momentum
is displayed, by taking $L =4M > L_{crit}$. These curves have a local
maximum (unstable circular orbit) and a local minimum (stable circular
orbit). The role of these two critical points
tends to interchange as
$\epsilon$ crosses $1/2$, as one
begins to push the limits of this perturbative calculation.
At still higher values of $L$, the character
of the maxima and minima become pronounced, as shown in Fig. 4, where
$L = 2\sqrt{10}M$. When the potential turns over, the minimum can
lead to the existence of bound orbits.

Unlike the null case, the impact parameter for massive particles depends
on the particle's angular momentum $L$. The capture cross section
$\sigma_{capture} = \pi b^2_{crit}$, where the critical impact
parameter is
$$b_{crit} = {{L_{crit}} \over {(E^2(r_{max}) - m^2)^{1/2}}},\eqno(34)$$
and $m$ denotes the test particle rest mass (in units where $c = 1$).
The energy $E$ in (34) is the amount required to just overcome the
potential barrier; i.e., $E(r_{max}) = V(r_{max})$, where
$r_{max}$ locates the maximum value of the effective potential.
The value of the critical angular momentum, defined to be that value
of $L$ below which there are no bound orbits, depends very weakly on
$\epsilon$, so that one may take $L_{crit} \approx 2\sqrt{3}M$ in
(34). However, as we will find later, $L_{crit}$ will depend
strongly on $N_v$, the number of gauge bosons, in the large-$N_v$
limit. With the exception of the gauge boson case, the capture cross
section for massive test particles tends to increase with the strength
of the back-reaction. This is due to the lowering of the potential
barrier at $r = r_{max}$.

\noindent
{\bf IV. Repulsive Gravity}

Additional physical insight into the consequences of the back-reaction
is provided by a study of the acceleration of a test particle in
the modified spacetime geometry (19). The acceleration gives a direct probe
of the force acting on the particle.

Here, we consider a massive test particle initially at rest; this is
equivalent to setting $L = 0$ in the timelike $(\kappa = 1)$
effective potential (30).
The four-velocity of a particle at rest is ($\tau$ denotes proper
time)
$$U^{\mu} = ({{dt} \over {d\tau}}, 0, 0, 0),\eqno(35)$$
and its acceleration
$$ a^{\mu} = {{dU^{\mu}} \over {d\tau}} + \Gamma^{\mu}_{\alpha \beta}
U^{\alpha} U^{\beta}.\eqno(36)$$
{}From (26) and (35),
with $\kappa = 1$ we have $(dt/d\tau)^2 = (- g_{tt})^{-1}$, so
the radial component of the acceleration is
$$a^{r} = \left({{dt} \over {d\tau}} \right)^2\, \Gamma^r_{tt} =
 {1 \over 2} g^{rr}\, {{\partial} \over {\partial r}}
\ln (-g_{tt}).\eqno(37)$$
Transforming to the particle's proper rest frame gives
$$a^{\hat r} =  {1 \over 2}  \left( g^{rr} \right)^{1/2} \,
 {{\partial} \over {\partial r}}
 \ln (-g_{tt}), \eqno(38)$$
where the caret refers to
components with respect to this frame;
{\it i.e.}, $g_{{\hat \mu} {\hat \nu}} = {\rm diag} (-1,1,1,1)$.

Evaluation of $a^{\hat r}$ requires knowledge of the metric components
$g^{rr}$ and $g_{tt}$ to $O(\epsilon)$. These can be read off from (19),
and are (when written in terms of $w$)
$$g^{rr} = (1 - w) \left[1 - \epsilon\,(1 - w)^{-1} w\, \mu (w) \right],
\eqno(39)$$
and
$$-g_{tt} = (1 - w) \left[ 1 + \epsilon \, (2 {\bar \rho}(w) -
  w (1 - w)^{-1} \mu (w)) \right].\eqno(40)$$
After some algebra, using (8) and (10) to calculate
${\partial \mu}/{\partial w}$ and ${\partial \rho}/{\partial w}$, we
arrive at the following compact expression
$$a^{\hat r} = (1 - w)^{-1/2}\, \left( {{w^2} \over {4M}} \right)
 \left[1 + \epsilon \, \Delta^{\hat r} \right],\eqno(41)$$
for the radial acceleration of a particle at rest, where
$$\Delta^{\hat r} \equiv (1 - w)^{-1} (1 - {{w} \over {2}}) \mu (w)
+ {{32\pi M^2} \over {\epsilon \, w^3}} \, T^r_r,\eqno(42)$$
is the correction due to the back-reaction.
$\Delta^{\hat r}$ is independent of $\epsilon$: indeed, the
``$\epsilon$-pole" in the second term is exactly cancelled by the
coefficient (4) which multiplies all the expressions for the stress-energy
tensors, (2).

It is a straightforward exercise to compute
$\Delta^{\hat r}$ for the scalar, spin-${1 \over 2}$ and vector boson
back-reactions by substituting the corresponding $\mu_j$ and
$T^r_{r,j}$ into (42) for $j = s, f$ and $v$.
We find that both
$$ \Delta^{\hat r}_s > 0 \qquad {\rm and} \qquad \Delta^{\hat r}_f > 0,
\, \qquad \forall r \geq 2M, \eqno(43)$$
but
$$\Delta^{\hat r}_v < 0,\eqno(44)$$
for $2.8M \stackrel{<}{\sim} r \stackrel{<}{\sim} 8.6M$.
In other words, while the conformal scalar and massless spinor
back-reactions appear to make the ``dressed" black hole more
attractive (we have seen this effect from the effective potential
point-of-view, in the lowering of the magnitude of the
effective potential for timelike orbits;
see Figures 2-4), the gauge boson back-reaction tends to
weaken the the attractive force of the black hole by generating a
localized spherical region or ``shell"
containing a repulsive component of the net
force. Since we have set $L = 0$
from the outset, this cannot be an artifact of the particle's orbital
motion (it is stationary), but must be ascribed
to a genuine repulsive gravitational
force, or {\it anti-gravity}, completely quantum mechanical in origin and
induced by the U(1) back-reaction.

\noindent
{\bf V. Multiple Field Back-reaction}

Thus far we have investigated the separate back-reactions due to
single species of conformal scalars, massless spinors and U(1) gauge
fields. While this has revealed novel important features of the back-reaction
problem, a more realistic setting should take into account black holes
in thermal equilibrium with
a heat bath comprising multiple species
of quantum fields. We know from elementary particle physics that the
replication or multiplicity of scalars, fermions and gauge fields
reflects the variety of quantum
numbers needed to distinguish physical
attributes (flavor, color, mass, etc.) observed directly or inferred
from observation.
Particle
replication is also the starting point for constructing unified models
of the fundamental interactions
based on large (i.e., rank 4 or greater) gauge groups [13].

Apart from the details of their specific phenomenologies, what
primarily distinguishes say, the standard model (SM) from one or more
of the grand unified theories (in which the SM must be embedded)
is the particle content and group-theoretic assignment
of each model. This is determined once the
choice of gauge group is made and the scalars (Higgs bosons) and
fermions (quarks and leptons) have been assigned to various
multiplets or representations of the group. These
group theory assignments can be
characterized with a set of integers. The number $N_v$
of gauge bosons
belonging to a given gauge group given by the number of group
generators [8,14].
Typically, the number of fermions $(N_f)$ is the
dimension of the representation times the number of families, and
the number of scalars ($N_s$) is determined by the pattern of symmetry
breaking to smaller groups one wishes to explore. We shall give examples
of the $N_j$
below for the standard model as well as for some
generic extensions of this model.

In the limit of free field theories on a background spacetime, the
back-reaction due to a collection of fields is easy to treat. This
follows since the stress-energy tensors in this limit only
depend quadratically on
the fields and are ``flavor-diagonal". To make this point clear, we
note the stress-energy tensor for a set of $N_s$ real massless conformally
coupled scalars is
$${\cal T}^{\mu}_{\nu} = \sum_{k = 1}^{N_s} \, (T^{\mu}_{\nu})^k,\eqno(45)$$
where (no sum on $k$)
$$(T_{\mu \nu})^k = {2 \over 3} \Phi^k_{,\mu} \Phi^k_{,\nu}
- {1 \over 6} g_{\mu \nu} g^{\sigma \rho} \Phi^k_{,\sigma} \Phi^k_{,\rho}
 - {1 \over 3} \Phi^k \Phi^k_{,\mu \nu} - {1 \over {12}}
g_{\mu \nu} \Phi^k \Box \Phi^k,\eqno(46)$$
is the tensor for a single scalar. Then, upon renormalization,
$$<{\cal T}^{\mu}_{\nu}>_{ren} = \sum_{k =1}^{N_s} <(T^{\mu}_{\nu})^k>_{ren}
 = N_s \, <T^{\mu}_{\nu}>_{ren},\eqno(47)$$
where the last equality follows from the fact that the renormalization
procedure is independent of the species label $k$.
Similar considerations hold for the renormalization of the spin-${1 \over 2}$
and gauge boson tensors in the multiple particle case [14].

The upshot of this is that we should obtain a good approximation to the
multiple field (and multiple spin) back-reaction in the limit of small
gauge, Yukawa and scalar
self-couplings by simply replacing the source term
in (1) by an appropriate weighted combination (with weights $N_s, N_f$
and $N_v$) of the single-species stress-energy tensors. From (1) and
(8) and (10), this entails the rescaling
$$\mu(r)_j \rightarrow N_j \, \mu(r)_j, \qquad {\rm and}
 \qquad \rho(r)_j \rightarrow N_j \, \rho(r)_j, \eqno(48)$$
for $j = s,f,v$.
To get a feeling for the values of the ``weights" one can expect, we
give a brief listing of the numbers of gauge fields, spinors and real
scalars contained in typical gauge theories in Table I.
The standard model of elementary particles contains a total of
12 gauge bosons (8 gluons, the $W^{+},W^{-},Z^{o}$ and the photon), requires
at least one complex scalar doublet (4 real scalars) for spontaneous
symmetry breaking, and has 3
families of 15 quarks and leptons (45 spinors) [15].
The Bose-Fermi symmetry of supersymmetry doubles the particle spectrum
of the non-supersymmetric models. For the minimal supersymmetric standard
model (MSSM), there are the 12 gauge bosons, the the spin-${1 \over 2}$
sector is augmented over that of the SM by the addition of 12 gauginos
(fermionic partners of the gauge bosons) and from the fermionic partners
of the two complex scalar doublets (8 real scalars) needed to break the
gauge symmetry and provide fermion masses
in this model [13]. This yields a total $N_f = 65$. The $45$
original fermions of the SM have 45 spin-0 partners which add to the
two complex doublets to give a total $N_s = 53$. The smallest
simple group
containing the SM is $SU(5)$. This has {\bf 24} gauge fields, 45 quarks and
leptons (3 families of 15), and scalars in the adjoint ${\bf 24}$
and a fundamental ${\bf 5}$ complex representation, or 34 real
scalars in total [13].
Finally, Table I includes an $E_6$ model which contains
{\bf 78} gauge particles,
three families of {\bf 27} fermions and
an adjoint {\bf 78} (real) and two {\bf 27}'s of (complex) scalars.

An important consequence of the multiple-field back-reaction
is that the validity of the perturbation theory breaks down
at a smaller radius, for a given $\epsilon > 0$, than for the
$N = 1$ cases. This can be appreciated immediately by inspection
of (21) and (22), which show that the metric perturbations grow
in direct proportion to the $N_j$. This growth shrinks the
domain-of-validity radius according to
$$1 \leq \left( {{r_{dom}} \over {2M}} \right)^2 =
 {{3K} \over {2 \alpha_j \, N_j}} \left( {{\delta} \over {\epsilon}}
\right). \eqno(49)$$
Nevertheless, because $K$ is such a large constant, we can still find
perturbatively valid solutions of (1) involving large numbers
($N_j >> 1$) of quantum fields. We illustrate this for the case of
the gauge boson back-reaction. In Fig. 5, we present a calculation of the
null potential for $N_v = 100$ and $\epsilon = 0.,\, .15,\, .3,\, .45$
and $.6$.
{}From (49), and taking $\delta = \epsilon$
for convenience, we find $r_{dom} = 27M$, which sets the scale for the
perturbation. Unlike any of the $N_j = 1$ cases discussed above, the
increased number of gauge fields prevent the barrier from turning
over with the rest of the potential as $\epsilon$ is increased.
An even more dramatic example is provided by taking $N_v =300$
($r_{dom} = 15M$), as shown in Fig. 6. Here, as the back-reaction
is increased, the barrier peak {\it increases} while the rest of the
potential flattens out. This phenomenon shows up only for the gauge
field case, and is due to the amplification of the repulsive
anti-gravity region which exists only for this case.
Note the no-back-reaction curve ($\epsilon = 0$) has the lowest barrier
of the set.
(We point out that the string-inspired unification group
$E_8 \otimes E_8$ contains 496 gauge bosons, so a
number like $N_v = 300$ is not unrealistic).

Similar amplification takes place for the timelike effective
potential, as shown in Fig. 7. Again, for illustrative purposes, we
take $N_v = 300$, the same range of $\epsilon$ as before,
but we set $L = 0$. Without the back-reaction, the
test particle would be doomed to be captured by the black hole, as is
well known. However, in the present case, we see the amplification of the
repulsive core may prevent the test particle from being captured--even
for a vanishing impact parameter!

\noindent
{\bf VI Discussion}

The lowest order solutions of the back-reaction problem solved in
this paper contain rather striking features which we have been
able to uncover through calculations of the black hole effective
potential. We have found the back-reaction tends to diminish the
overall magnitude of the potentials corresponding to null and timelike
orbits. This lowering of the effective potential is correlated with
an associated increase in the black hole capture cross sections for those
instances when the potential exhibits a barrier peak. This in turn will
affect the black hole lifetime which results from the competition between
particle capture and evaporation if the hole is formed in thermal
equilibrium and subsequently goes out of equilibrium with the
surrounding particle heat bath
[12]. For ``extreme" values of the
perturbation parameter ($\epsilon \stackrel{>}{\sim} 1/2$)
the potentials turn over completely, become negative near the
renormalized event horizon, and exhibit minima corresponding to bound
orbits. Although an $\epsilon$ approaching such
values is surely pushing the
limits of perturbation theory, perhaps beyond the bounds of even
qualitative reliability,
nevertheless, one may
interpret these results as indicating possible qualitative trends which
should be investigated in more nearly complete treatments.

The shapes and magnitudes of the effective potentials are similar for
the single-species back-reactions from
the spin-$0, {1 \over 2}$ and $1$ fields, but the U(1) case merits
special attention. The gauge boson back-reaction generates a repulsive
force in the neighborhood of the event horizon, which is revealed by
calculating the radial acceleration of a massive test particle
placed initially at rest outside the black hole. The component of the
net force due to the back-reaction points away from the origin, unlike the
scalar and fermion cases, where it points inward. The appearance of such
Casimir-type forces, which are quantum-mechanical in origin, should be
expected on general grounds. Indeed, as emphasized in [16], a meaningful
definition of the physical vacuum energy must take into account the
fact that quantum fields always exist in the presence of external
constraints, i.e., either in interaction with matter or other external
fields or boundaries. For the cases at hand, the renormalization of the
stress-energy tensors employed here must take into account that the
quantum fields (the scalar, spinor and gauge boson) interact with the
classical background Schwarzschild spacetime. This external ``constraint"
affects the zero-point modes of the quantum fields which in turn affects
the zero-point energies.

The treatment of multiple-species back-reaction is useful for problems
involving black holes in thermal equilibrium with heat baths made up
from fields belonging to representations of gauge theories. Again, the
gauge-boson example is particularly noteworthy since an increase in the
number of gauge fields can substantially amplify the repulsive
gravitational Casimir force to such an extent that the black hole
capture cross section
actually decreases relative to the no-back-reaction
limit. Since all spins discussed in this paper come into play for
the back-reaction due to a gauge theory
with matter, it may prove worthwhile
to study the cosmology of models that lead
to a net increase or decrease
in the capture cross section [12].

Provided the semiclassical back-reaction program leads qualitatively in the
right direction (that is, towards a correct quantum gravity), one should
include the spin-2 graviton contribution to the renormalized one-loop
effective stress-energy tensor. The effects of linear gravitons should
contribute a term to the stress-energy tensor of the same order as those
coming from ordinary matter and radiation fields.

Finally, not all scalars will
be conformally coupled to the curvature nor
will they necessarily be massless. For example, the Higgs scalars could
couple with any strength to the curvature, and the axion, if it
exists, couples minimally. Partial results from a calculation of the
renormalized stress-energy tensor of a scalar with arbitrary
coupling and mass has recently been published and could serve as
a starting point for a more general investigation of the scalar field
back-reaction [17].

\noindent
{\bf Acknowledgments}
This research was supported by DOE Grant No. DE-FGO5-85ER40226
(D.H. and T.W.K.) and by National Science Foundation grant
PHY-8908741 (J.W.Y.).\\

\noindent
{\bf References}

\noindent
[1]. M.A. Abramowicz and A.R. Prasanna, Mon. Not. R. Astr. Soc. {\bf 245},
720 (1990); M.A. Abramowicz and J.C. Miller, {\it ibid.} {\bf 245}, 729
(1990); M.A. Abramowicz, {\it ibid.} {\bf 245}, 733 (1990).

\noindent
[2]. K.W. Howard, Phys. Rev. D{\bf 30}, 2532 (1984); K.W. Howard and
P. Candelas, Phys. Rev. Lett. {\bf 53}, 403 (1984).

\noindent
[3]. B.P. Jensen and A. Ottewill, Phys. Rev. D{\bf 39}, 1130 (1989).

\noindent
[4]. D.N. Page, Phys. Rev. D{\bf 25}, 1499 (1982).

\noindent
[5]. M.R. Brown, A.C. Ottewill and D.N. Page, Phys. Rev. D{\bf 33},
2840 (1986).

\noindent
[6]. J.W. York, Phys. Rev. D{\bf 31}, 775 (1985).

\noindent
[7]. D. Hochberg, T.W. Kephart and J.W. York, Phys. Rev. D{\bf 48}, 479 (1993).

\noindent
[8]. D. Hochberg and T.W. Kephart, Phys. Rev. D{\bf 47}, 1465 (1993);
erratum to appear.

\noindent
[9]. T. Elster, J. Phys. A: Math. Gen. {\bf 16}, 989 (1983).

\noindent
[10]. T. Elster, Class. Quantum Grav. {\bf 1}, 43 (1984).

\noindent
[11]. C.W. Misner, K.S. Thorne and J.A. Wheeler, {\it Gravitation}
(Freeman, San Francisco, 1973). The effective potential is discussed in
Chap. 25.

\noindent
[12]. These results will be discussed in a forthcoming paper in progress.

\noindent
[13]. G.G. Ross, {\it Grand Unified Theories},
(Benjamin-Cummings, Menlo Park, 1984).

\noindent
[14]. In the abelian limit, the relation between
renormalized stress-energy tensors
for gauge groups $G$ and U(1) is $<T_{\mu \nu}>_{G} =
N_v \, <T_{\mu \nu}>_{U(1)}$. $N_v$ is given by the number of group
generators in $G$. In [8], the second Casimir invariant $C_2(G)$ should
be replaced everywhere by $N_v$.

\noindent
[15]. Chris Quigg, {\it Gauge Theories of the Strong, Weak, and
Electromagnetic Interactions}, (Benjamin-Cummings, Menlo Park, 1983).

\noindent
[16]. G. Plunien, B. M\"uller and W. Greiner, Phys. Rep. {\bf 134}, 87
(1986).

\noindent
[17]. P. Anderson, W. Hiscock and D.A. Samuel, Phys. Rev. Lett. {\bf 70},
1739 (1993).

\vfill\eject
\mediumtext
\begin{table}
\begin{tabular}{ccccc}
Model & Gauge Group & $N_v$ & $N_f$ & $N_s$ \\ \hline
Standard Model (SM) & $SU(3)\otimes SU(2) \otimes U(1)$  & 12 & 45 & 4 \\
Minimal SUSY SM  & $SU(3)\otimes SU(2) \otimes U(1)$ & 12 & 65 & 53 \\
Minimal SU(5) & $SU(5)$ & 24 & 45 & 34 \\
Three Family $E_6$ & $E_6$ & 78 & 81 & 186 \\
\end{tabular}
\caption{Particle multiplicities of various gauge theories}
\end{table}
\begin{center}
FIGURE CAPTIONS
\end{center}

\noindent
Fig. 1: Effective potential for null orbits: conformal scalar field
back-reaction. $\epsilon = 0.0, \, .15, \, .3, \, .45, \, .6$.

\noindent
Fig. 2: Effective potential, timelike orbits for $L = 0$: fermion
back-reaction.

\noindent
Fig. 3: Effective potential, timelike orbits
for $L = 4M$: fermion back-reaction.

\noindent
Fig. 4: Effective potential, timelike orbits
for $L = 2\sqrt{10}M$: fermion back-reaction.

\noindent
Fig. 5: Effective potential, null orbits: gauge
field back-reaction with $N_v = 100$.

\noindent
Fig. 6: Effective potential, null orbits: gauge
field back-reaction with $N_v = 300$.

\noindent
Fig 7:  Effective potential, timelike orbits
for $L = 0$: gauge field back-reaction with
$N_v = 300$.

\end{document}